%% file: main.tex
\documentclass[a4paper,11pt]{article}
\usepackage{pos}

\title{
The origin of MeV gamma-ray diffuse emission from the inner Galactic region
}
\ShortTitle{MeV gamma-ray diffuse emission and all sky}

\input{mypreset}

\usepackage[subrefformat=parens]{subcaption}
\author*[a,b]{Naomi Tsuji}
\author[c,b,d]{Yoshiyuki Inoue}
\author[e]{Hiroki Yoneda}
\author[f]{Reshmi Mukherjee}
\author[g]{Hirokazu Odaka}

\affiliation[a]{Faculty of Science, Kanagawa University, \\
    2946 Tsuchiya, Hiratsuka-shi, Kanagawa 259-1293, Japan}
\affiliation[b]{Interdisciplinary Theoretical \& Mathematical Science Program (iTHEMS), RIKEN \\
    2-1 Hirosawa, Wako, Saitama 351-0198, Japan}
\affiliation[c]{Department of Earth and Space Science, Graduate School of Science, Osaka University, \\
    Toyonaka, Osaka 560-0043, Japan}
\affiliation[d]{Kavli Institute for the Physics and Mathematics of the Universe (WPI), UTIAS, The University of Tokyo,\\
    5-1-5 Kashiwanoha,Kashiwa, Chiba 277-8583, Japan}
\affiliation[e]{Nishina Center, RIKEN, \\
    2-1 Hirosawa, Wako, Saitama 351-0198, Japan}
\affiliation[f]{Department of Physics and Astronomy, Barnard College, Columbia University, \\
    New York, NY, 10027, USA}
\affiliation[g]{Department of Physics, The University of Tokyo, \\
    7-3-1 Hongo, Bunkyo, Tokyo 113-0033, Japan}

\emailAdd{ntsuji@kanagawa-u.ac.jp}

\abstract{
The origin of the inner Galactic emission, measured by \comptel\ with a flux of $\sim ~ 10^{-2}$~\Mflux\ in the 1--30 MeV range, 
has remained unsettled since its discovery in 1994. 
We investigate the origin of this emission by taking into account individual sources which are not resolved by \comptel\ and the Galactic diffuse emission. 
The source contribution is estimated for sources crossmatched between the \bat\ and \lat\ catalogs by interpolating the energy spectra in the hard X-ray and GeV gamma-ray ranges, as well as unmatched sources. This results in a flux of $\sim$20\% of the \comptel\ excess. 
The Galactic diffuse emission is calculated by \galprop\ to reconcile the cosmic-ray and gamma-ray spectra with observations by AMS-02, \voyager, and \lat, resulting in a flux of $\sim$30--80\% of the \comptel\ emission. 
Thus, we show that the COMPTEL emission could be roughly reproduced by a combination of the sources and the Galactic diffuse emission. 
Furthermore, combined with the extragalactic emission,
we construct all-sky images in the MeV gamma-ray range
to pinpoint some potential interesting targets for future missions, 
which would be critical for bridging the ``MeV gap'' in the spectra of gamma-ray sources.
}

\FullConference{%
  7th Heidelberg International Symposium on High-Energy Gamma-Ray Astronomy (Gamma2022)\\
  4-8 July 2022\\
  Barcelona, Spain
}

\begin{document}
\maketitle

\section{Introduction} 
\label{sec:intro}

The MeV gamma-ray domain is the only unexplored window among recent multiwavelength observations in astrophysics, often referred to as the ``MeV gap''.
One of the open issues in MeV gamma-ray astrophysics is the origin of diffuse emission from the inner Galactic region.
%
The Imaging Compton Telescope \comptel\ onboard the Compton Gamma-Ray Observatory (CGRO) reported the detection of diffuse emission of 10$^{-2}$ \Mflux\ in 1--30 MeV from the inner Galactic region with $|\ell| \leq 30\degr$ and $|b| \leq 15\degr$ \citep{bouchet_diffuse_2011,strong_diffuse_1996}.
This emission was derived by considering the instrumental background and \ac{cgb}, thus it would contain \ac{gde} and unresolved sources.
Recently, the inner Galactic diffuse emission has been confirmed by other observations such as \spi\ \cite{siegert_diffuse_2022} and the electron-tracking Compton camera (ETCC) aboard the balloon mission of SMILE-2$+$ \cite{takada_first_2022}.

The origin of the inner Galactic diffuse emission has been in active debate. 
If the \ac{gde} model to account for the diffuse emission by \lat\ \cite{ackermann_fermi_2012} is extrapolated to the MeV energy range, 
there is an apparent excess component (e.g., \cite{strong_diffuse_2004}), which is commonly referred to as the \comptel\ excess.
There are several scenarios for reproducing the \comptel\ excess:
(1) Individual MeV gamma-ray sources should be taken into consideration. 
(2) There are non-negligible uncertainties on the model of \ac{gde}, since it has a lot of unconstrained parameters (e.g., photon field densities, \ac{cr} source distribution, \ac{cr} injection spectra, and propagation mechanism).
Enhancement of one or more of these parameters can make \ac{gde} higher to reach the \comptel\ excess \citep{bouchet_diffuse_2011}.
(3) New populations, such as annihilation or decay of dark matter \citep{boddy_indirect_2015} and/or cascaded gamma rays accompanying cosmic neutrinos \citep{fang_tev_2022}, might be present.

In this proceeding, we investigate the \comptel\ excess by taking into account MeV gamma-ray sources unresolved by \comptel\ and \ac{gde} in \secref{sec:analysis}. We also explore all-sky images in the MeV gamma-ray range in \secref{sec:allsky}. 

\section{
The inner Galactic diffuse emission in MeV}
\label{sec:analysis}

\subsection{MeV gamma-ray sources}
\label{sec:sources}

Although the previous studies (e.g., \cite{strong_diffuse_1996,orlando_imprints_2018,siegert_diffuse_2022}) proposed that the \comptel\ excess would be attributed by radiation from individual unresolved sources,
the quantitative estimation has not been done yet.
We estimate this source contribution from a MeV gamma-ray source catalog in \cite{tsuji_cross-match_2021}, which presented a crossmatching between the 105-month \bat\ \citep{Bird2016} and 10-yr \lat\ (4FGL-DR2) \cite{4fgldr2} catalogs, resulting in 156 point-like and 31 extended crossmatched sources.
These crossmatched sources, which are both hard X-ray and GeV gamma-ray emitters, are prominent sources in the MeV gamma-ray sky.
%
Among them,
19 point sources and 14 extended sources are located in the inner Galactic region with $|\ell| \leq 30\degr$ and $|b| \leq 15\degr$.

We jointly fit the \acp{sed} of \bat\ in 14--195 keV and \lat\ in 50 MeV--300 GeV.
For the fitting model, we adopt log-parabola or two-component models, depending on the sources. The two-component model is a superposition of the models in the \bat\ and \lat\ catalogs.
%
Based on the best-fit model 
determined by the minimum $\chi^2$, we estimate spectra in the MeV gamma-ray energy range, sum up all the spectra of the 33 sources in the inner Galactic region, and divide it by the region size.
The result of the accumulated source spectrum is shown in \figref{fig:model}.
The contribution of all crossmatched sources to the \comptel\ excess is about 10\%. 
See \cite{tsuji_2022} for details.

Besides the crossmatched sources in \cite{tsuji_cross-match_2021}, there exist many unmatched sources that would have a significant contribution accumulatively.
In the region with $|\ell| \leq 30\degr$ and $|b| \leq 15\degr$, there are 152 \bat\ and 708 \lat\ sources, where the crossmatched sources are excluded.
These unmatched sources (860 in total) would be fainter than $10^{-12}$ \flux\ in the MeV energy band since they are not detected by \bat\ or \lat, with the sensitivity being approximately 10$^{-12}$~\flux.
If we assume each unmatched source has a flux of  $10^{-12}$ \flux,
the accumulative source flux is $\sim 10^{-3}$ \Mflux, 
which should be considered as an upper limit.
This upper limit flux of the unmatched sources is roughly comparable to that of the crossmatched sources.
Combined with the crossmatched sources, the contribution of the sources is $\sim$20\% of the \comptel\ excess (\figref{fig:model}).

\subsection{Galactic diffuse emission}
\label{sec:gde}

To evaluate \acf{gde}, we make use of \galprop\ (version 54 of WebRun), which is designed to calculate astrophysics of \acp{cr} (i.e., propagation and energy loss) and photon emissions in the radio to gamma-ray energy bands \citep{porter_high-energy_2017}.
In this proceeding, we use the \ac{gde} models in the literature \cite{ackermann_fermi_2012,orlando_imprints_2018}, which are developed to be consistent with the \ac{cr} observations by AMS-02 (and \voyager) and the gamma-ray observations by \lat.
A baseline model of 
$^{\rm S}S ^{\rm Z}4 ^{\rm R}20 ^{\rm T}150 ^{\rm C}5$\footnote{This model assumes that the source distribution of \acp{cr} is SNRs, the Galactic disk is characterized by the height of $z=$4 kpc and the galactocentric radius of $R=$20 kpc, and $T_s$=150 K and $E(B-V) = 5 $ mag cut is adopted for determining the gas-to-dust ratio (see \cite{ackermann_fermi_2012} for details).} 
is selected as a representative of the models in \cite{ackermann_fermi_2012} and referred to as Model~1. 
From \cite{orlando_imprints_2018}, we adopt the DRE (i.e., diffusion and re-acceleration) and DRELowV (modified DRE\footnote{
DRE model with some modifications on parameters of diffusion and particle injection in order to reproduce the \ac{cr} measurements \cite{orlando_imprints_2018}.}) models, hereafter referred to as Model 2 and Model~3, respectively.

Since GDE below $\sim$100 MeV is dominated by the \ac{ic} component, the difference in Models 1--3 arises from \ac{cr} electrons in 0.1--1 GeV.
Models 2 and 3 are respectively the highest and lowest with the difference of a factor of a few, and Model 1 is in the middle of them.
Although there is such uncertainty on the \ac{gde} models, $\gtrsim$30\% of the \comptel\ excess is contributed by GDE.

\subsection{Results and discussion}
\label{sec:discussion}

\begin{figure}[ht!]
\centering
\includegraphics[width=0.9\hsize]{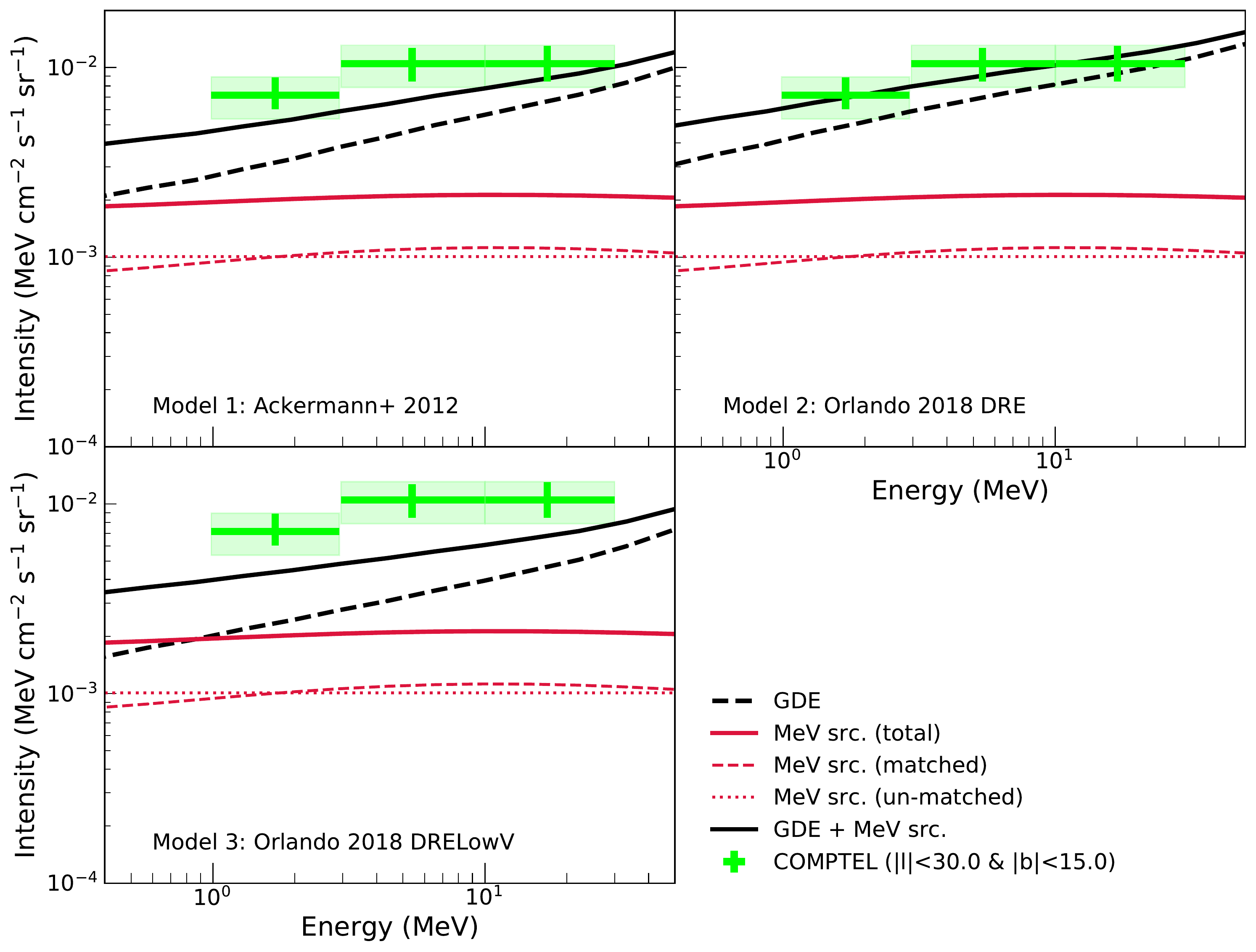}
\caption{
The \acp{sed} of \ac{gde} (dashed black line) and
the sources (solid red line for all the sources, dashed red for the crossmatched sources, and dotted red for the unmatched sources).
The combined spectrum of these two components is illustrated with a solid black line.
The results with \ac{gde} Models 1, 2, and 3 are respectively shown in the upper left, upper right, and lower left.
The \comptel\ emission \citep{bouchet_diffuse_2011} is indicated by light green points, 
and light green squares are its systematic error \citep{strong_diffuse_1994}.
\label{fig:model}}
\end{figure}

\figref{fig:model} compares the \comptel\ data points and our models with the three different models of \ac{gde}.
We show the spectrum of MeV gamma-ray sources, \ac{gde} Models~1--3, and the combined spectra of these two components.
It should be noted that this direct comparison did not take into account energy dispersion, which may have a significant effect on the result, as described in the data analysis of \spi\ \citep{strong_gamma-ray_2005}.
%
We find that the combination of \ac{gde} and the sources can roughly reproduce the \comptel\ excess:
the entire spectrum can be sufficiently explained with Model~2 (\figref{fig:model} upper right), and
the lowest and highest energy bins of the \comptel\ data can be reproduced with Model~1 (\figref{fig:model} upper left).
Model~3 (\figref{fig:model} lower left) is slightly lower than the \comptel\ excess.

The spectral shape would provide us with a new constraint.
The power-law spectral index is $s \sim -1.9$ ($dN/dE \propto E^s$) for the \comptel\ excess.
\figref{fig:model} shows that the spectrum of the accumulated sources is almost flat in the SED with a spectral index of $s \sim -2$.
Since the \ac{gde} models have $s \sim -1.5$, the MeV gamma-ray sources should play an important role in reproducing the observed spectrum by \comptel\ with $s \sim -1.9$.
More precise measurements of the spectrum of each source will enable constraining the accumulative source spectrum, which in turn will be useful to determine the \ac{gde} spectrum, especially the index of the primary electrons responsible for the \ac{ic} radiation.

Since the dominant component in the energy range of the \comptel\ excess is \ac{gde}, the uncertainty of \ac{gde} prevents us from reaching a robust conclusion.
The uncertainty arises from the amount of \ac{cr} electrons.
The CR electrons in 100--1000 MeV, which produce 1--30 MeV photons via \ac{ic} scattering, are different by a factor of $\sim$4, depending on the models. 
To distinguish these models is important in the perspective of \ac{cr} feedback on galaxy evolution:
\acp{cr} can produce a non-thermal pressure gradient and enhance the degree of ionization in molecular clouds, significantly affecting a star-forming activity
(e.g., \cite{hopkins_first_2021}). 
The density of the photon fields across the Galaxy is not well constrained, which also makes the \ac{gde} model somewhat uncertain in the MeV gamma-ray band.
If we assume that the interstellar radiation field is locally enhanced, such as in the Galactic bulge or a large Galactic CR halo, \ac{gde} in the inner Galactic region is increased enough to reach the flux level of the \comptel\ excess \citep{bouchet_diffuse_2011}.

As indicated by Model~3, we would need additional component(s) to reconcile with the \comptel\ emission.
Possible explanations are low-mass ($\lesssim$280 MeV) annihilating dark matter coupling to first-generation quarks \citep{boddy_indirect_2015} and/or 
cascaded gamma rays accompanying cosmic neutrinos \citep{fang_tev_2022},
which would open up a new window for these studies.

\section{MeV gamma-ray all-sky map} \label{sec:allsky}

\begin{figure}[t!]
\centering
  \begin{minipage}[b]{0.45\linewidth}
    \centering
    \includegraphics[keepaspectratio, width=\hsize]{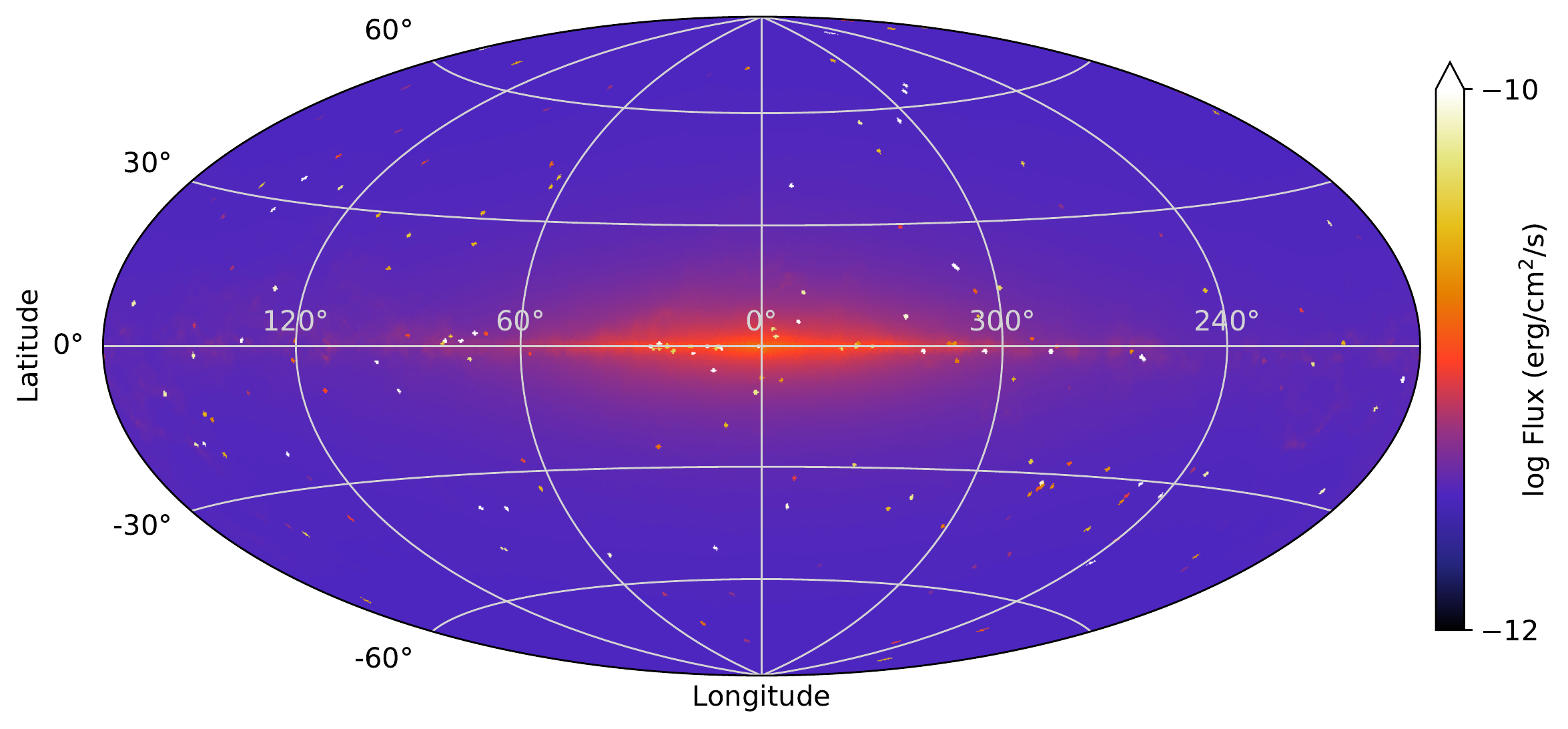}
    \subcaption{All components}
  \end{minipage}
  \begin{minipage}[b]{0.45\linewidth}
    \centering
    \includegraphics[keepaspectratio, width=\hsize]{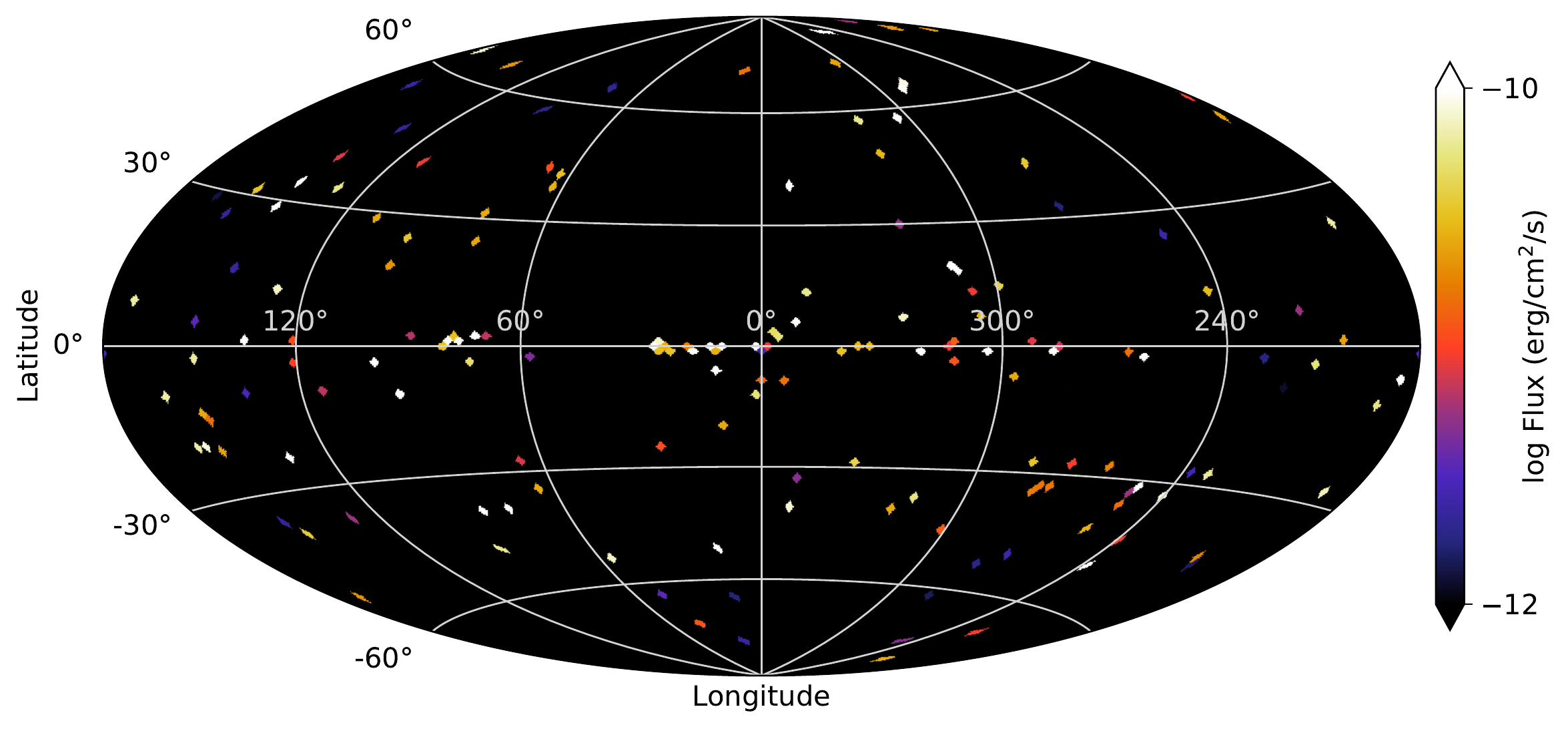}
    \subcaption{Source map}
  \end{minipage}
  \vspace{1mm}
  \begin{minipage}[b]{0.45\linewidth}
    \centering
    \includegraphics[keepaspectratio, width=\hsize]{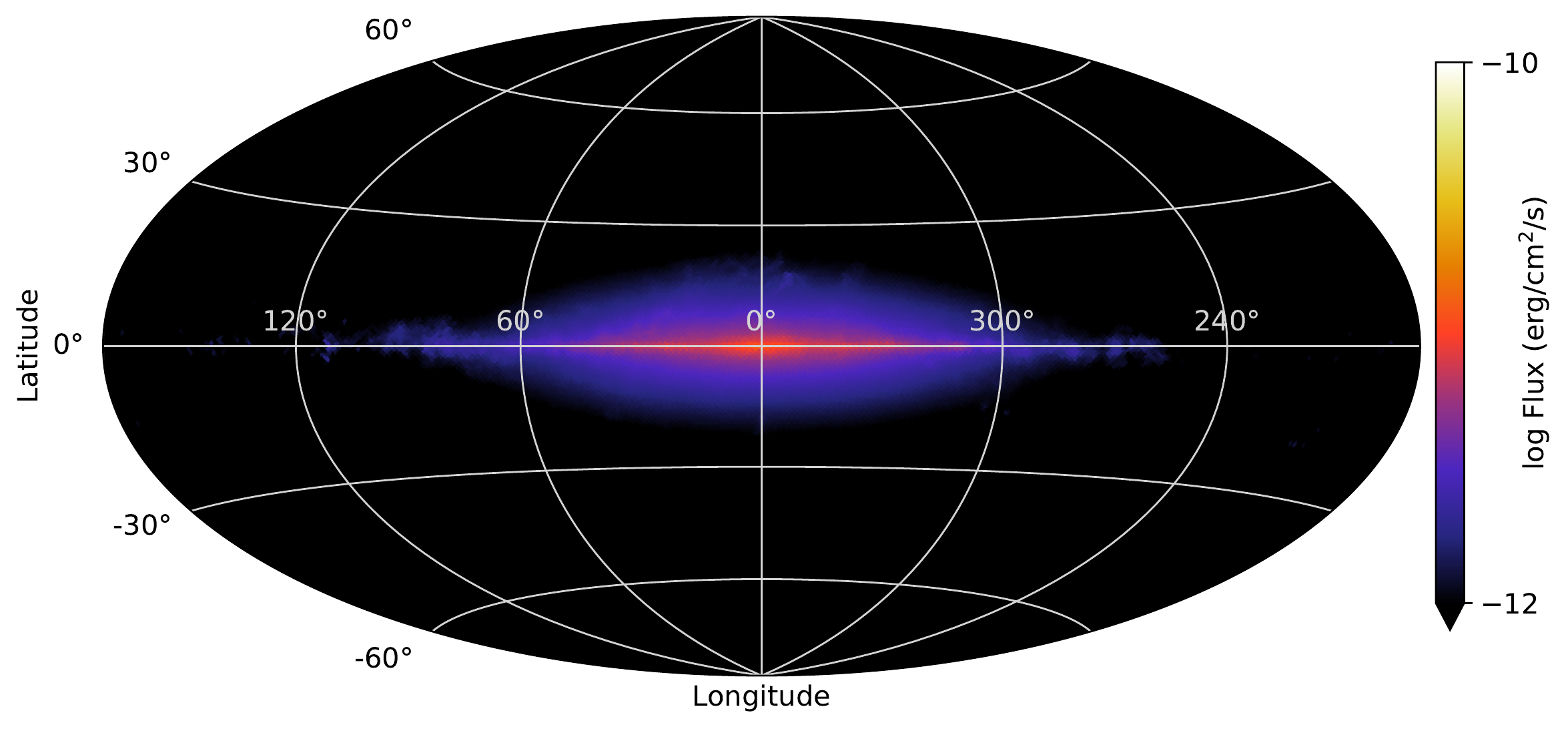}
    \subcaption{\acf{gde} Model~1}
  \end{minipage}
  \begin{minipage}[b]{0.45\linewidth}
    \centering
    \includegraphics[keepaspectratio, width=\hsize]{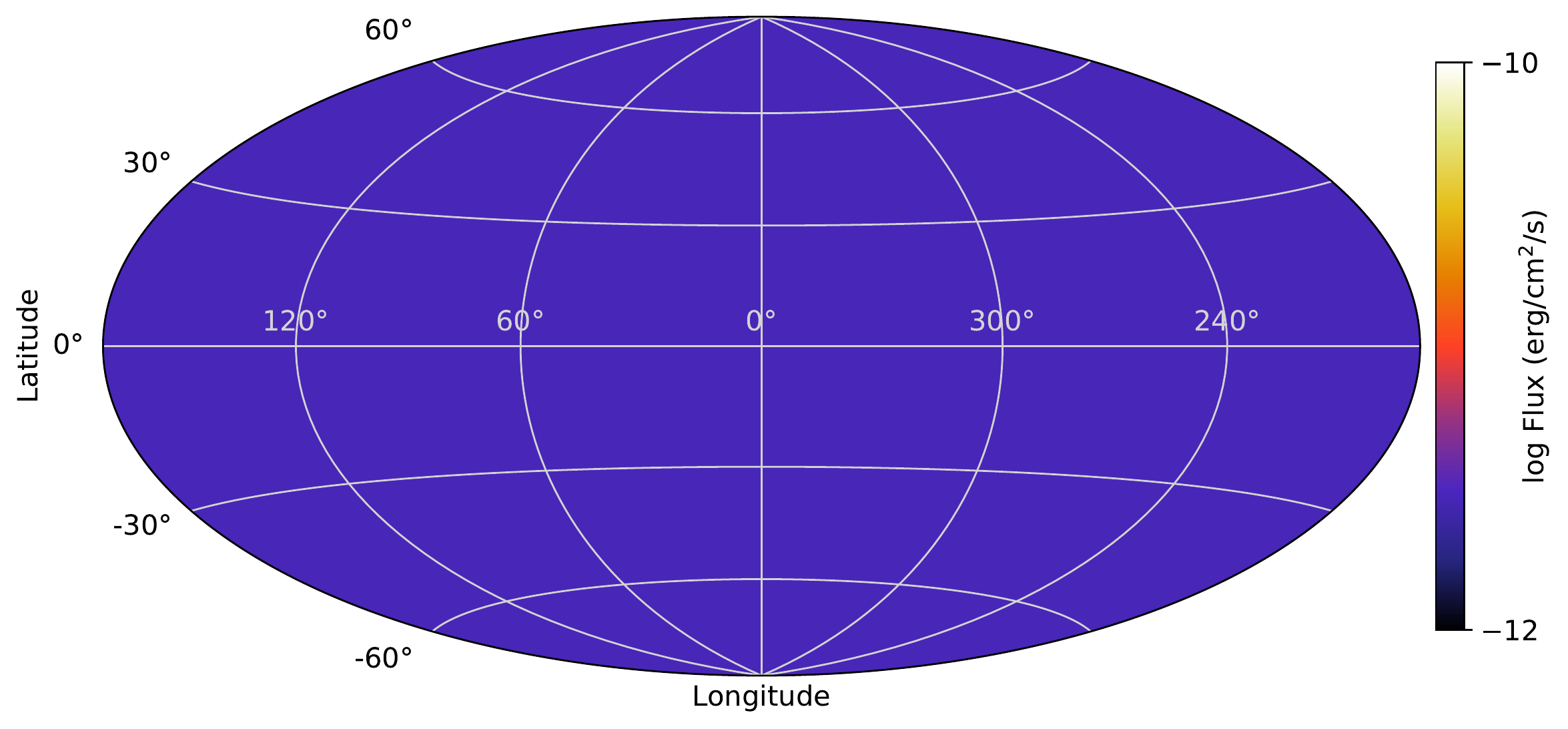}
    \subcaption{Cosmic gamma-ray background (CGB)}
  \end{minipage}
\caption{
The all-sky maps of the 1--10 MeV flux, shown in the Galactic coordinate and hammer projection. 
\label{fig:allsky}
}
\end{figure}

This section presents predicted MeV gamma-ray all-sky maps, which would be useful for observation strategies of future missions.
The all-sky map comprises three components: sources, \acf{gde}, and \acf{cgb}.
The source map (\figref{fig:allsky}b) consists of 187 crossmatched sources in \cite{tsuji_cross-match_2021}. 
The flux is estimated by fitting the \bat\ and \lat\ \acp{sed} by log-parabola or tow-component models, as mentioned in \secref{sec:sources}.
The extended sources are illustrated as point sources here, but this will be improved in the future work.
The \ac{gde} map (\figref{fig:allsky}c) makes use of Model~1 \cite{ackermann_fermi_2012} described in \secref{sec:gde}.
For the \ac{cgb} map (\figref{fig:allsky}d), we assume the spectrum in the literature \cite{weidenspointner_cosmic_2000} and the isotropic distribution.

The 1--10 MeV all-sky maps of each component and the total are illustrated in \figref{fig:allsky}.
\ac{gde} is dominant in the low-latitude sky, while it is dominated by \ac{cgb} at the higher latitude.
The flux of \ac{gde} and \ac{cgb} are respectively 0.18--15 and 3.0 in units of $10^{-12}$~\flux,
and 147 sources have flux larger than these values at each location.
Our model predicts that there are 123 and 174 sources with the flux in 1--10 MeV being larger than $10^{-11}$ and $10^{-12}$ \flux, respectively.
These sources would be good targets for future missions \cite{COSI_SMEX,fleischhack_amego-x_2021,Aramaki2020,takada_first_2022}.


\paragraph{Data release}
We provide all of the resources of this study (MeV gamma-ray source catalog, Galdef files of the \ac{gde} models, and all-sky FITS files) in \url{https://tsuji703.github.io/MeV-All-Sky}, which will be updated sometimes.

\acknowledgments
We thank the GRAMS collaboration \cite{Aramaki2020} and the MeV gamma-ray community.
N.T. acknowledges support from the Japan Society for the Promotion of Science KAKENHI grant No. 22K14064.




\input{references_selected.tex}


\end{document}

%% file: mypreset.tex
\def\figref#1{Figure~\ref{#1}} 
\def\eqref#1{Equation~\ref{#1}} 
\def\secref#1{Section~\ref{#1}} 

\def\hessj1731{HESS J1731$-$347}

\def\bat{{\it Swift}-BAT}

\def\integral{{\it INTEGRAL}}

\def\spi{{\it INTEGRAL}-SPI}

\def\comptel{COMPTEL}

\def\lat{{\it Fermi}-LAT}

\def\voyager{\textit{Voyager}}

\def\galprop{{\sc GalProp}}

\def\flux{\hbox{${\rm erg}~{\rm cm}^{-2}~{\rm s}^{-1}$}}
\def\Mflux{\hbox{MeV~cm$^{-2}$~s$^{-1}$~sr$^{-1}$}}

\def\pizero{\hbox{$\pi^0$}}

\usepackage[]{acro}

\DeclareAcronym{gde}{
  short = GDE ,
  long  = Galactic Diffuse Emission,
}

\DeclareAcronym{cgb}{
  short = CGB ,
  long  = cosmic gamma-ray background,
}

\DeclareAcronym{grb}{
  short = GRB ,
  long  = gamma-ray burst,
}
 
\DeclareAcronym{agn}{
  short = AGN ,
  long  = active galactic nucleus ,
  long-plural-form = active galactic nuclei
}
\DeclareAcronym{fsrq}{
  short = FSRQ ,
  long  = flat spectrum radio quasar ,
}
  
\DeclareAcronym{lmxb}{
  short = LMXB ,
  long  = low mass X-ray binary ,
  long-plural-form = low mass X-ray binaries ,
}
\DeclareAcronym{hmxb}{
  short = HMXB ,
  long  = high mass X-ray binary ,
  long-plural-form = high mass X-ray binaries ,
}
  
\DeclareAcronym{ic}{
  short = IC ,
  long  = inverse Compton ,
}

\DeclareAcronym{nustar}{
  short = {\it NuSTAR} ,
  long  = Nuclear Spectroscopic Telescope Array ,
}

\DeclareAcronym{bi}{
  short = BI ,
  long  = backside illumination ,
}
\DeclareAcronym{fi}{
  short = FI ,
  long  = frontside illumination ,
}
\DeclareAcronym{fov}{
  short = FoV ,
  long  = field of view ,
}
  
\DeclareAcronym{sim}{
  short = SIM ,
  long  = Science Instrument Module ,
}
\DeclareAcronym{hetg}{
  short = HETG ,
  long  = High Energy Transmission Grating ,
}
\DeclareAcronym{letg}{
  short = LETG ,
  long  = Low Energy Transmission Grating ,
}
\DeclareAcronym{hrc}{
  short = HRC ,
  long  = High Resolution Camera ,
}
\DeclareAcronym{acis}{
  short = ACIS ,
  long  = Advanced CCD Imaging Spectrometer ,
}
\DeclareAcronym{hrma}{
  short = HRMA ,
  long  = High Resolution Mirror Assembly,
}

\DeclareAcronym{compton}{
  short = Compton ,
  long  = Compton Gamma Ray Observatory,
}
\DeclareAcronym{hst}{
  short = HST ,
  long  = Hubble Space Telescope ,
}
\DeclareAcronym{iact}{
  short = IACT ,
  long  = Imaging Atmospheric Cherenkov Telescope ,
}
\DeclareAcronym{wcd}{
  short = WCD ,
  long  = Water Cherenkov Detector ,
}

\DeclareAcronym{hawc}{
  short = HAWC ,
  long  = High-Altitude Water Cherenkov ,
}
\DeclareAcronym{cta}{
  short = CTA ,
  long  = Cherenkov Telescope Array ,
}

\DeclareAcronym{em}{
  short = EM ,
  long  = electromagnetic ,
}
\DeclareAcronym{ism}{
  short = ISM ,
  long  = interstellar medium ,
}
\DeclareAcronym{csm}{
  short = CSM ,
  long  = circumstellar medium ,
}
\DeclareAcronym{sne}{
  short = SNe ,
  long  = supernovae , 
}

\DeclareAcronym{iss}{
  short = ISS ,
  long  = International Space Station ,
}

\DeclareAcronym{uhecr}{
  short = UHECRs ,
  long  = ultra high energy cosmic rays , 
}
\DeclareAcronym{ta}{
  short = TA ,
  long  = Telescope Array , 
}
\DeclareAcronym{auger}{
  short = Auger ,
  long  = Pierre Auger Observatory , 
}
\DeclareAcronym{ams}{
  short = AMS ,
  long  = Alpha Magnetic Spectrometer , 
}
\DeclareAcronym{pamela}{
  short = PAMELA ,
  long  = Payload for Antimatter Matter Exploration and Light-nuclei Astrophysics , 
}
   
\DeclareAcronym{cmb}{
  short = CMB ,
  long  = Cosmic Microwave Background , 
}
\DeclareAcronym{sed}{
  short = SED ,
  long  = spectral energy distribution , 
}
\DeclareAcronym{mhd}{
  short = MHD ,
  long  = magnetohydrodynamical ,
}
\DeclareAcronym{dof}{
  short = dof ,
  long  = degree of freedom ,
}
\DeclareAcronym{cco}{
  short = CCO ,
  long  = central compact object ,
  first-style = default
}
\DeclareAcronym{lmc}{
  short = LMC ,
  long  = Large Magellanic Cloud ,
}
\DeclareAcronym{smc}{
  short = SMC ,
  long  = Small Magellanic Cloud ,
}

\DeclareAcronym{hess}{
  short = H.E.S.S. ,
  long  = High Energy Spectroscopic System ,
  first-style = default
}
\DeclareAcronym{snr}{
  short = SNR ,
  long  = supernova remnant ,
}
\DeclareAcronym{pwn}{
  short = PWN ,
  short-plural = e ,
  long  = pulsar wind nebula ,
  long-plural  = e ,
}

\DeclareAcronym{sn}{
  short = SN ,
  short-plural = e ,
  long  = supernova ,
  long-plural  = e ,
  first-style = default
}

\DeclareAcronym{nw}{
  short = NW ,
  long  = northwest ,
  first-style = default
}
\DeclareAcronym{hxc}{
  short = HXC ,
  long  = hard X-ray component ,
  first-style = default
}

\DeclareAcronym{cr}{
  short = CR ,
  long  = cosmic ray ,
}
\DeclareAcronym{psf}{
  short = PSF ,
  long  = point spread function ,
}
\DeclareAcronym{hpd}{
  short = HPD ,
  long  = half power diameter ,
}
\DeclareAcronym{fwhm}{
  short = FWHM ,
  long  = full width of half maximum ,
}

\DeclareAcronym{pic}{
  short = PIC ,
  long  = particle-in-cell ,
  tag = numerical ,
}
\DeclareAcronym{cxb}{
  short = CXB ,
  long  = Cosmic X-ray Background ,
}
\DeclareAcronym{grxe}{
  short = GRXE ,
  long  = Galactic Ridge X-ray Emission ,
}
\DeclareAcronym{pa}{
  short = PA ,
  long  = Positional Angle ,
}
\DeclareAcronym{dsa}{
  short = DSA ,
  long  = diffusive shock acceleration ,
}

\def\pizero{\hbox{$\pi^0$}}

\def\degr{\hbox{$^\circ$}}

\def\utw{\smash{\rlap{\lower5pt\hbox{$\sim$}}}}
\def\udtw{\smash{\rlap{\lower6pt\hbox{$\approx$}}}}
